\documentclass{article}

\usepackage{PRIMEarxiv}

\usepackage[utf8]{inputenc} 
\usepackage[T1]{fontenc}    
\usepackage{hyperref}       
\usepackage{url}            
\usepackage{booktabs}       
\usepackage{amsfonts}       
\usepackage{nicefrac}       
\usepackage{microtype}      
\usepackage{lipsum}
\usepackage{fancyhdr}       
\usepackage{graphicx}       
\graphicspath{{media/}}     

\title{Investigating the Software Engineering Roadmap for Smart City Infrastructure Development: Goals and Challenges} 

\author{
  Mamdouh Alenezi \\
  Software Engineering and Disruptive Innovation (SEDI) \\
  College of Computer and Information Sciences \\
  Prince Sultan University \\
  Riyadh, Saudi Arabia\\
  \texttt{malenezi@psu.edu.sa}
}

\begin{document}
\maketitle

\begin{abstract}
In today's world, many cities are embracing cutting-edge technology and transforming into "smart cities". These emerging innovations are revolutionizing the standard of living for people, and as a result, smart city infrastructure development has become a major focus for city planners and policymakers worldwide. The goal is to create more livable, sustainable, and efficient urban environments, and software engineering plays a crucial role in achieving this. In this article, we will delve into what makes a city "smart" and what it means for the future. We will explore the software engineering roadmap for smart city infrastructure development, highlighting the goals and challenges that come with this innovative approach to urban planning. Our aim is to provide valuable insights into the importance of software engineering in achieving successful smart city infrastructure development. 
As cities continue to grow and evolve, it is essential to adopt new technologies that can help us build smarter, more sustainable communities. Smart city initiatives are paving the way for a brighter future, and software engineering is at the forefront of this movement. By understanding the software engineering roadmap for smart city infrastructure development, we can work towards creating more livable, efficient, and sustainable urban environments for generations to come.
\end{abstract}


\section{Introduction}

Any city that employs digital technologies to improve the quality of life for its residents is considered a smart city \cite{anthopoulos2017rise}. Information and communication technologies, which redefine urban services like transportation, connectivity, and energy, among others, are thus at the center of the city’s regeneration. So, by integrating those technologies into its existing infrastructure, any existing city may become smarter. The objective is to build an intelligent network of interconnected machines that communicate with one another in real-time and enhance the quality of life for every resident \cite{mohanty2016everything,jasim2021design,ismagilova2019smart,ansari2018store}. Due to population increase, major cities all over the world are experiencing congestion and a breakdown of their public services. Due to its viability from an environmental, social, and economic standpoint, the process of transforming those cities into smart cities seems to be the best answer to the problems brought on by the growth of conglomerates \cite{moya2017impacts,moyano2021traffic}.

\begin{figure}[htbp]
\centering
\fbox{\includegraphics[width=.92\linewidth]{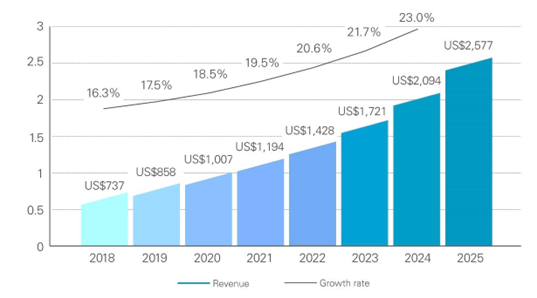}}
\caption{Global smart city market (US \$ billion), 2018-2025 (Source: PwC and “Smart Cities Market Analysis \& Segment Forecasts to 2025,” Grand View Research, 2018}
\label{fig1}
\end{figure}

The word ”smart city,” which is a catch-all for how technology, as well as advanced analytics, could be used to improve sustainability and efficiency in cities, can make a significant impact on how residents and decision-makers redefine urban areas. A smart city strategy has long been regarded as an essential tool for controlling cities’ exploding population increase \cite{brettle2020accelerating}. And over half of the globe’s population, which could increase to almost 68 percent by 2050, currently resides in urban areas, as per United Nations Department for Economic and Social Affairs. Cities’ financial influence has increased along with their population expansion. As per McKinsey \cite{kacyira2016urbanization,ritchie2018urbanization}, the leading 600 cities in the world are predicted to produce 60\% of the world’s gross domestic product by 2025. Figure \ref{fig1} shows the global smart city market in US \$ billions from 2018 to 2025 according to Cities Market
Analysis \& Segment Forecasts done by PricewaterhouseCoopers (PwC). 

It has been hard to construct the latest smart cities from the ground up compared to upgrading existing urban facilities. The sophistication of such projects and their huge price tags have decelerated deployment. Even so, the effects of COVID-19 on financial systems, societies, as well as public health could still inspire cities to conquer such obstacles and find innovative, multidisciplinary solutions to make smart city standards the cornerstone of all everlasting urban development plans. The global epidemic might also offer beneficial information and backing for the application of particular technological innovations \cite{brettle2020accelerating}.

An urban area that utilizes different electronic data collection sensors to convey content used to handle assets and resources effectively is referred to as a ”smart city” in particular \cite{camero2019smart}. The term "smart cities" can refer to a variety of things, though, depending on who is using it \cite{gil2015smarter,belli2020iot}. Those responsible for managing government services, municipal government, democratic accountability, environmental balance, as well as ICT systems, for example, may view smart cities from very different angles. The solutions that a smart city provides attract a lot of interest from its citizens.

Despite popular belief, the idea of smart cities is not new. Today’s digital technology developments, such as big data, machine intelligence, gadget interconnection, as well as rising consumer expectations, are what gave rise to the discussion about smart cities. Connectivity of application domains that generate the immersive areas that smart cities guarantee to be at the center of any smart city’s smart base, which is necessary for smart cities. The demands of consumers have transformed. The smart city progression aims to create hyper-connected environments by integrating advanced, data, as well as the latest electronic technologies.

Software Engineering is the application of a disciplined and systematic, which is a computable approach, to the development, operation, and maintenance of software \cite{larrucea2017software}. It studies the software development lifecycle from the beginning to the end. Software Engineering can produce a big line of software systems. These software systems can be products, platforms, product lines, etc \cite{fahmideh2021software}. Software engineering processes and procedures form the quality management core to develop and maintain software systems cost-effectively \cite{zarour2021software}. The development of smart systems requires new tools and technologies, as with any other developing paradigm in software engineering, but also an awareness of how the development cycle works to help software engineers implement these systems in a reliable and organized manner.

Software engineering is a crucial component of the modern smart city, as it allows cities to leverage technology to better serve their citizens. By developing applications, software engineers can ensure that urban systems are connected and functioning optimally while providing citizens with access to the latest digital services and tools. As cities become more connected, software engineers will be needed to design, develop, and maintain the necessary systems that will drive the smart cities of the future. In addition, software engineers will be essential in helping cities become more sustainable and efficient, as well as developing innovative applications to help cities become more livable, vibrant, and efficient.

According to Sonata’s Platformation™ process, software platforms must possess these characteristics as Open, Scalable, Connected, and Intelligent. These characteristics must be applied to any future platforms that are created. We can classify the current ecosystem as a platform if it displays these traits. A platform is an ecosystem of resources and a collection of software that enables you to expand your business. A platform’s value derives not only from its features but also from its capacity to link external tools, teams, data, and processes \cite{jansen2013defining,chen2018microservices,bogers2019ecosystem}. Now, the digital infrastructure that will enable tomorrow's smart cities is being developed, in part because to software development. The sophistication of our cities and the advancement of societal life will depend on intelligent technologies that can help us in many facets of our daily lives. Since the things that software development produces will directly affect how we live in our cities and towns, it follows that software development itself must become more intelligent. Several experiments are currently being tested all across the world. Several experiments are currently being tested all across the world. Although these are solitary initiatives centered on certain technologies (driverless vehicles, face recognition for security, and e-learning platforms in schools), the time when they all come together in smarter cities is not far off.

Thousands of distributed smart objects, platforms, backbone services, and software platforms and applications may be a part of smart city systems, which are networked and interconnected to offer value-added services and intelligent reasoning to other systems and end users \cite{larrucea2017software}. This class of systems’ design and implementation is increasingly becoming the norm for smart cities, retail, logistics, manufacturing, agriculture, transportation, healthcare, and other applications \cite{fahmideh2020exploration}. Basic services provided by the underlying software infrastructure are necessary for integrating various domains into a comprehensive and consistent solution. Such services, including facilities for application development, integration, deployment, and management to enable the construction of complex Smart Cities applications, could be offered via a revolutionary, all-encompassing software platform.

Because the architecture of smart cities will be based on a combination of potent technologies, software development will be crucial. It will be extremely difficult to develop platforms and technologies that can support the millions of devices that make up the "neural network" of a smart city since they will need to integrate cutting-edge trends like artificial intelligence, machine learning, cloud computing, and automation. In addition, development teams will need to consider the big picture because the applications they design will affect millions of people's lives in terms of society and the environment.

Platform engineering principles largely apply to DevOps because its fundamental ideas revolve around change management flexibility, automation, continuous delivery, and quick feedback response. With its best practices and technologies, platform engineering is made possible by Sonata’s platform engineering blueprint, which pro- motes continuous innovation, automation, greater productivity, and shortened time to market \cite{thomas2014architectural,senapathi2018devops,zarour2019research,hein2020digital}. Engineers built cities with hard materials such as steel and concrete. However, software engineers need solid stable architecture, flexible APIs, and proper management tools to build modern software that ties everything together. Smarter software development is the key to producing linked apps that will turn the concepts of smart cities into intelligent, physical realities.

The main objective of this paper is to investigate a software engineering roadmap for smart city infrastructure. Section 2 discusses major conceptual initiatives proposed in the literature, focusing on the primary consideration of each process and decision. Section 3 presents the research methodology steps. Section 4 presents a few top software engineering goals for smart cities which must be met substantially. Section 5 discusses the top smart city software engineering challenges. Section 6 provides a summary of the fundamental elements of smart cities and some recommendations, as well as Section 7, provides concluding statements and future work suggestions. 

\section{Literature Review}

Industry 4.0, Smart Cities, Internet of Things (IoT) are considered new technologies that are classified as contemporary software systems. Such technologies require a shift from the classical monolithic view of development to embedded interconnected smart systems and object ecosystems. Software engineering and development have to change from developing a monolithic structure to a multidisciplinary approach. Finding a secure, flexible, scalable, flexible, and cost-efficient architecture, platforms, and APIs are essential to managing and coping with the complexity of smart city software ecosystems \cite{motta2018challenges}.

Although there are numerous smart city initiatives underway in various nations around the globe \cite{neves2020impacts,hu2021smart,tomivcic2019smart}, these deployments are frequently built on custom systems that are neither interoperable nor portable across cities, expandable, or cost-effective \cite{rhee2018global}. The scientific community must address significant issues regarding software engineering in order to overcome these restrictions.  These difficulties call for highly specialized expertise and a thorough comprehension of various topics, but they also call for tight collaboration amongst experts from other professions who work together. From the perspective of software infrastructure, key research problems related to smart cities are urgently required. The creation of a comprehensive open-source platform with all the necessary components should receive more attention if strong, robust, integrated, and sophisticated applications are to be created for future smart cities. In order to promote the state-of-the-art in software engineering, this paper seeks to inspire scholars to conduct creative research on smart cities.

Du et al. \cite{du2018sensable} described in detail the architecture, applications, and future research directions of the smart city. da Silva et al. \cite{da2013smart} described numerous architectures while emphasizing the primary goals that each one tries to achieve. A list of specifications for the implementation of a smart city was also presented and analyzed, based on various configurations with the broadest range of reasons. Haque et al. \cite{haque2022conceptualizing} explored various aspects of the smart city literature to discuss architecture along with various applications along with notable implementations focusing on security and privacy issues.

Stages towards a prototype that could harmonize the viewpoints of the specialties in the Urban Systems Collaborative were described by Harrison \& Donnelly \cite{harrison2011theory}. They initiated by offering instances of smart cities as well as explaining why this progression is so innovative. They described how information technology shapes new behavioral norms meant to support the maintenance of high populations. The important premise of the Urban Systems Collaborative was therefore explained, namely that as information becomes more widely available, researchers also created urban systems models which can aid citizens, business owners, community groups, and governments in understanding more regarding how their cities function, how individuals use it, how they experience about it, where the city has issues, and what kind of solutions can be used.

A reference structure for Smart City development that could be used as the layout language for specifications for such cities was described by Abu-Matar \cite{abu2016towards}. They believed that a plan like that would support various stakeholders, gadgets, portals, as well as innovations. They contend that such extremely broad and diverse environments involved a novel design strategy. Therefore, they offered a reference architecture that draws inspiration from various software engineering initiatives like service-oriented architecture (SOA). Rather than having to start from scratch, the suggested reference architecture could very well act as a blueprint as well as a be- ginning point that includes contemporary architectural components, best practices, and trends. They presented a preliminary meta-model with different perspectives that emphasize relationships between individual views. They also reviewed the current state of the art, as well as offering a research roadmap for the entire smart city software development community.

The basic design principles of a smart city were briefly presented by Khatoun \& Zeadally \cite{khatoun2017cybersecurity}, who also reviewed current smart city projects as well as initiatives. They first discussed different privacy as well as security challenges, suggestions, as well as guidelines for smart city areas and their facilities before recognizing numerous security flaws and privacy concerns related to smart cities that should be discussed. Nikolov et al. \cite{nikolov2016learning} addressed a summary of a conceptual framework and offered designs for the design and implementation of intelligent learning settings. The introduction focuses on an assessment of emerging companies and new job categories that will require future workers to be well-equipped to address the requirements of such industries’ growth requirements as well as stay consistent with their infrastructure requirements.

According to Gottschalk \& Uslar \cite{gottschalk2015supporting}, smart cities could be endorsed by connecting various smart areas, like smart grids as well as smart homes, to a single, sizable area. Information systems are used to link things together, and they are supported by clearly defined functionalities as well as interactions. According to them, it is challenging to present a comprehensive picture of the functionalities of smart cities and other smart areas at this time because they are still in the development stage. Because of this, the two methodologies—use case methodology as well as integration profile were presented in their work and also implemented by a web-based application. From the perspective of smart cities, Farias et al. \cite{farias2019designing} looked into the construction process for smartphone platforms. They conducted multi-case research with 9 applications from 4 various developmental organizations to create a grounded theory to confront the dearth of credible information regarding designing mobile applications. The architectural style of each application was revealed through reverse engineering of the applications. A preliminary grounded concept was postulated depending on all the information to describe how the chosen design process creates an app with the required characteristics. The observed theory provides justifications for how groups of software engineers create mobile applications for smart cities. According to them, with this understanding, the phenomenon would be better understood, as well as interpretations of the design and development phases would be more precise.

The development of smart city infrastructure is an important area of focus for many municipalities around the world. Smart city infrastructure relies heavily on software engineering to enable the integration of various technologies and systems that make cities more efficient, sustainable, and livable. The goals of smart city infrastructure development include improving the quality of life for citizens, increasing efficiency in city operations, reducing resource consumption, and enhancing public safety. To achieve these goals, smart city infrastructure development requires the use of various software engineering tools and techniques, including software design, development, testing, deployment, and maintenance.

However, developing smart city infrastructure also poses several challenges. One significant challenge is the complexity of the systems involved. Smart city infrastructure typically involves the integration of various systems, such as transportation, energy, water, and waste management systems. These systems often have different architectures, protocols, and data formats, making integration difficult. Another challenge is the need for interoperability. As smart city infrastructure involves the integration of various systems from different vendors, it is crucial to ensure that these systems can communicate with each other seamlessly. This requires the development of interoperable standards and protocols that can be adopted by different systems.

Security and privacy are also significant challenges in smart city infrastructure development. As smart city infrastructure involves the collection and sharing of large amounts of data, it is crucial to ensure that this data is protected from unauthorized access and misuse. Additionally, privacy concerns may arise if personal data is collected and shared without the consent of individuals. Overall, the development of smart city infrastructure presents significant opportunities and challenges for software engineering. To overcome these challenges, software engineers need to adopt a holistic approach to system design and development, which takes into account the unique requirements and challenges of smart city infrastructure.
\section{Research Methodology}

As stated in a literature review, in the coming years, we require a sensible city that makes utilizes the resources delivered by acquired data and communication technology to enhance our living standards. We can also say that in the coming decades if we want to live a comfortable lifestyle, we will be accustomed to ICT to make the most of limited resources such as a house, power, transportation, time, and so on. Today's world leaders strive to create in the areas of services and lifestyle quality by anticipating people's needs and utilizing technology to monitor the operations of environmental elements. Many systems employ sensors, a storage facility for information, as well as a system that specialists and analyses the data before making a call in accordance with that. A previous study on the development of technological advances, software products, as well as markets is used to assist firms in planning their technical development in various domains and as a strategy for technological forecasting. The majority of prior smart mapping approaches concentrated on enhancement and improvement. In response to fast industrial development and transformation, this study provides a three-phased well-organized and effective technology mapping strategy.  This research is based on standard smart city innovations, which were gathered for this article and utilized to assemble the core technologies as well as software systems that smart cities are currently usually adopting.

Previously, the procedure of technological development planning methodology was overly complicated, therefore product life cycles were constantly shortened and more diverse. Previous research on assessment to evaluate focused on the significance of products and technologies. Moreover, we feel that the market (service applicability) sector is a significant aspect of technology roadmaps because various applications, as well as products, can be deployed in various smart city fields. As a result, this study primarily focuses on the relationship between technology and applications, as well as we streamline the procedure to achieve an efficient decision-making mechanism. We don't highlight the physical product directly, but rather the future use of services.

Stage I is the initiation stage, during which the research inquiry, literature, as well as case selection are defined. Stage 2 is the technological inventory stage, which categorizes both technologies as well as applications based on current literature and previous reports. Stage III of technological maturity is the specialist prediction stage, in which technological advancement forecasts are employed to demonstrate the causal association between a technology and the date of its deployment. As a result, we may divide the software engineering road mapping process for smart city development (Fig. 2) into the steps outlined:

\begin{figure}[htbp]
\centering
\fbox{\includegraphics[width=.72\linewidth]{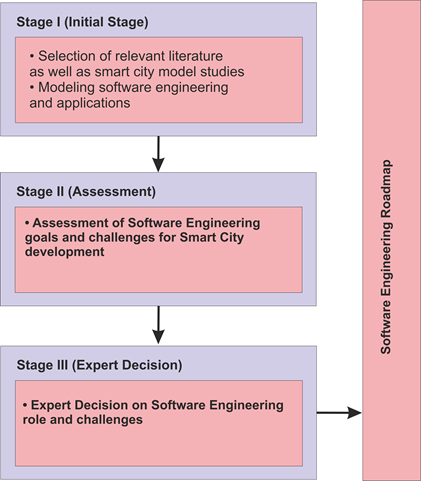}}
\caption{Software engineering road mapping process for smart city development}
\label{fig2}
\end{figure}

\section{Top Software Engineering Goals for Smart Cities}

Smart cities' software engineering has to aim for different critical goals if they are to succeed. These goals are: 

\subsection{Citizens’ lives Improvement}

Enhancing the quality of life for city residents is the main goal of deploying technology there. It could imply a variety of things. For example, it might entail developing a new public transportation system that makes use of autonomous cars that are powered by artificial intelligence and machine learning. For those who live in remote sections of the smart city, the combination of these two technologies may mean the elimination of traffic jams, improved routes with shorter travel times, and far more effective commutes \cite{macke2018smart,macke2019smart,zhu2022can}. There are plenty of additional things that technology may do to enhance city living. With the development of new approaches for smart cities, it will be possible to control traffic through automated devices that can react to accidents and traffic jams in real-time, assist teachers through e-learning platforms integrated into schools, improve water distribution, intelligent waste management with recycling at its core, and many other things.

\subsection{Human activity environmental impact reduction}

The enormous effect our activities are having on the environment is one of the main worries for contemporary society. There is no denying that climate change has increased the frequency and severity of natural disasters around the world, which have a significant impact on our cities. Through a number of initiatives, smart cities could help to lessen our influence on the environment.

For instance, sensors placed around the city might track the number of chemicals or pollution in the air and water, generating alerts for the more polluted areas and providing information on what can be done to lessen the problem (such as limiting the traffic, increasing recycling, installing new water treatments systems in the area, etc.) \cite{chen2018water,lee2020design,liu2021alleviating}. Moreover, sensors might be employed in parks and public areas to monitor the maintenance of trees and plants, preserving the natural environment in smart cities. These sensors might be a part of a larger automated system that monitors humidity and sunshine levels and sets off particular actions to maintain the health of the plants. This system might use artificial intelligence and machine learning. Making systems more energy-efficient and low energy use is another consideration.

\subsection{City’s safety Enhancement}

Due to recent discussions about facial recognition systems and potential privacy invasions, public safety control has gained popularity. Such discussions serve as just one illustration of how technology might be applied to enhance safety in public spaces in "smart cities"; they highlight both the advantages of the strategy as well as its implementation's difficulties.

The debate continues despite the fact that society has not reached a consensus regarding facial recognition in public areas. This is because its proponents believe that this technology could be used to both deter potential criminal behavior in controlled environments and identify criminals and terrorists in general. In crowded places like train and bus stations, facial recognition can also be used to locate kidnapped children, elderly people, and missing people.

In terms of security, software development can produce more than that for a smart city. The resident's safety may also be improved by road monitoring technologies and automated lighting systems. Numerous examples from around the world have demonstrated that monitoring public spaces could be of great assistance to the police \cite{wan2017smart,zywiolek2021perception,del2021inclusiveness}.

\subsection{Industrialized areas optimization}

Being close to industrial areas is one of the most unsightly things about living in a city. This is because they are frequently polluted, noisy, and crowded.  Because of these factors, surrounding neighborhoods are frequently neglected, which results in underdeveloped areas that are true stains in the fabric of cities. Additionally, a number of industries must dispose of undesirable by-products of their processes while also consuming a significant amount of energy and water.

Technology can help make all of that work better. Power stabilization systems and the use of renewable energy could both help make better use of electric energy. Special software could look at real-time data, look at how energy is used, and anticipate high demands from industries that need to be prepared and use energy efficiently.

In addition, technology has the potential to intelligently manage the supply of water in a manner analogous to that of energy management across a variety of industries. A smarter industry is developing as a result of the installation of smart air conditioning systems and waste treatment plants. Lastly, utilizing intelligent logistics systems may reduce traffic, maximizing fleet efficiency and minimizing its impact on the surrounding areas.

\subsection{Cities, regions, and countries collaboration}

The creation of a better world was always an underlying promise in all technological advancement. This is unquestionably a promise that smart cities hold, not only for their residents but also for the entire planet. This is due to the fact that smart software used in these cities could link up with similar applications used in other cities, regions, and even nations to accomplish the same objectives. The sharing of vital data and platforms may result in the growth of regions beyond the boundaries of the smart city, improving everyone's quality of life. There may be new business opportunities for a large portion of society, an increased limit on the environmental impact of human activity across nations, and increased security in larger regions. It is undeniable that smarter cities and the technological solutions they employ can result in a world that is more interconnected and has a higher standard of living.

\subsection{Accessibility of Infrastructure}

Ensuring that smart city infrastructure is accessible to all citizens is an essential aspect of smart city development. Software engineers can play a crucial role in making smart city infrastructure accessible to everyone by adopting the following approaches:

\begin{enumerate}
    
\item Designing for accessibility: Software engineers should design smart city infrastructure with accessibility in mind, incorporating features such as assistive technologies, user-friendly interfaces, and universal design principles. This can help to ensure that all citizens, including those with disabilities, can access and use smart city infrastructure.

\item Conducting user testing: Software engineers should conduct user testing with a diverse group of citizens, including those with different abilities and needs, to ensure that smart city infrastructure is accessible to all. User testing can help to identify accessibility issues and inform design decisions that can improve the accessibility of smart city infrastructure.

\item Providing training and support: Software engineers should provide training and support to citizens to help them effectively use smart city infrastructure. This can include providing documentation, tutorials, and other resources that can help citizens with different abilities and needs access and use smart city infrastructure.

\item Ensuring equitable access: Software engineers should ensure that smart city infrastructure is accessible to all citizens, regardless of their socioeconomic status or location. This can be achieved by deploying smart city infrastructure in areas that are underserved or marginalized, such as low-income neighborhoods or rural areas.

\item Adhering to accessibility standards: Software engineers should adhere to accessibility standards, such as the Web Content Accessibility Guidelines (WCAG), when designing and developing smart city infrastructure. Adhering to these standards can help to ensure that smart city infrastructure is accessible to all citizens, including those with disabilities.

\end{enumerate}
Software engineers can ensure that smart city infrastructure is accessible to all citizens by adopting a user-centered design approach, conducting user testing, providing training and support, ensuring equitable access, and adhering to accessibility standards. By doing so, software engineers can help to create inclusive and accessible smart cities that benefit everyone.
\section{Top Software Engineering Challenges for Smart Cities}

More and more cities adopt the concept of smart cities to manage their processes and satisfy the increasingly sophisticated demands of their citizens. Unfortunately, due to the complexity and lack of effective support by current infrastructures, developing smart-city systems is a very tedious task. The issues involved in developing software systems for smart cities are not fully addressed by the present technologies and methodology offered by current practices in software engineering [47]. In this scenario, thousands of individuals are creating systems that must be intricately integrated, leading to even more complicated systems of systems [48]. The degree of spread and heterogeneity of these systems as well as their magnitude, both in terms of users and machines, will rise explosively as a result of this trend. It will become more usual in this setting to build systems made up of people, services, and ”things” that interact with one another to carry out distributed tasks. Combining interoperability, mobility, heterogeneity, adaptability, privacy, and security issues to these large-scale systems will make them more complex to be deployed in future Smart Cities. We identify 6 categories of software engineering challenges in the context of smart-city systems and propose some solutions. These are, (1) developing the necessary models for smart-city systems; (2) system components management and optimization; (3) designing tools, methods, and models for critical infrastructures; (4) necessary quality attributes optimization; (5) integrating software; and (6) considering a high level of interoperability, evolvability, configurability, and adaptability is required while creating a smart infrastructure.

\subsection{Challenge 1: Model development for smart cities}

Developing models can be at the right level of abstraction where a high-level representation of the topic of interest and hiding unnecessary details of software so that complexity can be managed. These models can also enhance reusability and extensibility through model reuse and model transformations while enforcing correctness through model-based verification techniques and testing and can reduce development time by generating code from these models.

Model-Driven Engineering (MDE) is a well-known approach in Software Engineering that is based on models, meta-models ( and meta-meta models, etc.), and model transformations \cite{snoeck2022agile}. During the last decade, there has been an increasing emphasis on MDE, and as such many useful models are readily available for use. We strongly believe that the lack of models for designing smart-city applications is one of the causes of the excessive programming effort needed in realizing smart-city applications.

\subsection{Challenge 2: Framework design for managing sub-components}

Smart-city systems typically consist of systems of systems or ecosystems \cite{obaidat2016smart}. Each system is assumed to be a (potentially) independent software system interoperating with other systems to achieve common business goals. Both the elements of a system and the systems of systems may be required to be configured according to the needs. To create generic infrastructures, the infrastructure must provide means to integrate/operate a variant set of applications. Optimizing configuration requires modeling the common and variable aspects of these systems. Since smart cities have to cope with various sets of requirements, the design of configuration that satisfies the management and quality requirements is not trivial. We believe that managing and configuring the elements of clusters should not be restricted to procedural measures only (such as initializing, linking, etc.) but the overall quality of configurations must be considered as well. For system structuring, generic MDE frameworks have not been developed so far to manage and optimize the elements of systems or systems of systems according to the user-defined quality optimization criteria.

\subsection{Challenge 3: Models, methods, and tools design for critical infrastructures}

Critical infrastructure for smart cities can be defined as “an asset or system which is essential for the maintenance of vital societal functions within a city”. Damage, destruction, or disruption of critical infrastructure caused by natural disasters, terrorism, criminal acts, or malicious acts can have a significant negative impact on the safety of city residents \cite{coman2019vulnerability,parn2019cyber,heino2019critical}.  It should be noted that this term is typically used at the scope of a nation, but similarly, it applies also to smart cities which will heavily rely on the cyber part (software) that controls the city. We claim that specially designed software systems can effectively support designing critical infrastructures by providing architectures with a high degree of reliability, availability, security, timeliness, and correctness. Dependability is the ability of a system to deliver a service that can justifiably be trusted. This definition encompasses several quality attributes including reliability, availability, etc.

Reliability is defined as the continuity of correct service, whereas availability is readiness for correct service. Computer security or cybersecurity is defined as “the protection of computer systems from theft or damage to their hardware, software or electronic data, as well as from disruption or misdirection of the services they provide”. Timeliness is defined as the ability of a computer/software system to complete its processing within the specified time constraints. Correctness is defined as the fulfillment of the functional (and qualitative) requirements of the system. To address this challenge, it is considered necessary to design and implement various architectural styles for critical infrastructures within the context of an MDE framework. Adopt an MDE approach to define architectural styles for reliability and availability, security and timeliness; define quality models for reliability/availability, security and timeliness; define trade-off relationships among the quality attributes, and define models for multi-objective optimization criteria. Designing, implementing and evaluating innovative software architectures for dis- tributed, large-scale, self-configurable, and self-adapting systems based on smart algorithms and machine learning methods are among the challenges. Additionally, it should focus on creating new approaches, strategies, software engineering tools, and IDEs to assist in the creation, testing, and implementation of complex, large-scale networked systems of the highest quality for Smart Cities. It should make use of the technologies created as part of the project’s implementation in real-world scenarios involving actual smart city services, to enhance citizen quality of life and offer solutions for more effective management of big and megacities.

\subsection{Challenge 4: Quality attributes optimization through system adaptation at run-time}

Smart-city systems are long-living systems. When a system is deployed, it must be adapted, modified, extended and (partially) replaced while operational. This challenge deals with the automatic adaptation of smart-city infrastructures at run-time. There has been a large number of proposals to verify systems at run-time \cite{aktas2019provenance}. Most of these systems adopt various specification languages and generate run-time monitors and verification modules according to defined specifications. Although a considerable number of systems are now available for use, there have been only a few proposals to verify systems across multiple languages in distributed system settings. As such most existing run-time verification systems are less suitable for verifying systems of systems. Several approaches have been presented for designing and implementing so-called quality-aware architectures, for various purposes, for example for reducing energy consumption. Self-adaptive or self* systems have been defined in the literature to create systems with dynamic adaptation \cite{spinner2019online}.

Most of these systems are inspired by the adaptive control theory and as such, they adopt one or more control loops for adaptation. Recently, within the context of MDE approaches, several research proposals have been presented to design adaptive systems. So-called models@run-time systems adopt various feedback control mechanisms to improve the system performance according to the predefined control parameters. However, there is hardly any MDE framework available that allows user-defined quality models, monitors the systems accordingly at run-time to check if the desired quality attributes are satisfied, and optimize the system structure using self-adaptation mechanisms. Abdullahi et al. \cite{abdullahi2022software} conducted a systematic literature review for quality attributes of smart cities' software architecture. They adopted the “ISO/IEC 25010:2011” standard for quality attribute classification. Based on their review, four key quality attributes are found to be essential based on the frequency of appearance of each quality attribute. These attributes are security, portability, performance efficiency, and compatibility.

\subsection{Challenge 5: Software Systems Integration}

Integrating software systems can be practically realized at the system level or programming-language level. “System integration is defined as the process of bring- ing together the component sub-systems into one system and ensuring that the subsystems function together as a system. There are various integration possibilities such as horizontal and vertical integration, which roughly correspond to communication-channel-based integration or layered architecture-based integration, respectively. Programming language level integration is realized at a much finer level of language modules. A typical smart-city application may require the integration of many sub-systems and corresponding programming-language modules, for example, data sensing, gathering, and reasoning, run-time configuration, and tuning, storing data on cloud systems, analyzing data, adopting machine learning techniques, utilizing decision support systems, business processes, and work-flow schedulers, etc.

Many of the functionally specialized sub-systems and language modules/libraries may be developed independently. Integration is generally carried out using techniques like IDLs and stub-generators, adaptors, proxies, scripting languages, glue-code, brokers, virtual machines, dedicated libraries, internet protocols, web services, service-oriented architectures, etc. Although these techniques help considerably, they may still require considerable programming effort and in-depth knowledge about the systems/modules being integrated. The difficulties in using these techniques originate from the following observations:
\begin{itemize}
\item	Some of these techniques are imperative techniques meaning that the programmer is obliged to write a considerable amount of program code to realize the integration. Techniques such as adaptors, proxies, scripting languages, glue-code, dedicated libraries, etc. fall into this category.
\item	Some of the techniques are applicable only at the system level and mostly require dedicated system/vendor-specific solutions. Techniques such as IDLs and stub-generators, virtual machines, web services, service-oriented architectures, etc. fall into this category. Since systems evolve continuously, the borders of integration cannot be fixed; programming-language level integration may turn out to be a system-level integration, or vice versa, etc. System/vendor-specific integration may also be a limiting factor in the evolution of systems.
\end{itemize}

Software engineers want easy-to-integrate solutions that make smart cities easier to operate. For consumers, this means being able to complete everyday tasks in a friction-free environment. To ensure a seamless experience, smart cities must enable consumers to purchase goods and services with the speed and security they expect in many other environments. A connected platform that offers seamlessly integrated payment solutions is essential to providing a unique payment experience for every business in the smart city. Allowing consumers to confidently swipe, dip, and tap to complete transactions enables the on-the-go pace that so many smart cities offer.

\subsection{Challenge 6: A high degree of interoperability, evolvability, configurability, and adaptability in designing smart infrastructure}

These quality attributes ensure flexibility in software architecture. Flexible language models and architectural styles ease coping with changes both at compile time and run-time. If the underlying languages and architectural styles are too rigid, the design space of the alternatives of design and runtime adaptations is too limited. Since smart-city systems are long-living systems, extending the lifetime of the systems becomes then very difficult. The concepts of separation of concerns and composition of concerns are fundamental in designing languages with a high degree of flexibility. The motivation here is to reduce the complexity of software by decomposing software into manageable parts. Explicit composition operators are defined to create flexible systems. In practice, the term flexibility may refer to various quality attributes. Providing language mechanisms and architectural styles that fulfill the flexibility needs of smart-city applications has not been accomplished yet. Moreover, it is not practical to introduce a new language when so many languages are available for use. New language proposals, therefore, must extend the existing languages instead of offering completely new language semantics and syntax.
\section{Discussion}

Contrary to popular belief and despite the recent surge in media attention, the concept of smart cities is not a novel idea. The emergence of modern technology, such as big data, machine intelligence, and device connectivity, coupled with the increasing expectations of consumers, has ignited the conversation surrounding smart cities. At the core of any smart city lies a system of application domains that generate interactive spaces. However, the development of smart cities requires a solid foundation. The demands of consumers have evolved, and the objective of smart city development is to create a hyperconnected environment by integrating advanced technologies, utilizing data, and leveraging cutting-edge innovations.

To improve operational capabilities and meet the demands of people, numerous city governments across the world support the idea of smart cities. A smart city is described as an urban ecosystem that makes use of connected technologies to enhance city operations and enable wise decision-making in close to real-time. According to Gartner, a ”smart city” is an urbanized area where various sectors collaborate to achieve sustainable results through the interpretation of situational, reliable details shared between sector-specific data as well as operational technological tools. Such an environment consists of several disparate and widely dispersed cyber-physical components where various stakeholders must communicate and work together. A variety of subject matter experts, including decision-makers, engineering technicians, developers, and urban planners, are involved in designing the architectural features of a smart city. As a result, many obstacles prevent these stakeholders from collaborating and communicating with one another, including a lack of a prevalent design language, varied backgrounds, competing priorities, and a variety of development tools, to mention a few.

Software designers would want to have simple-to-integrate solutions which will make interacting with smart cities convenient. For customers, this involves having a frictionless surrounding in which to carry out daily tasks. Customers would need to be able to buy goods as well as assistance in smart cities with the reliability and safety they’ve grown to anticipate in several other settings if they want to have a seamless experience. To provide each user of a smart city with customized payment expertise, a linked system with seamless integration of payment processing is essential. Customers can conduct transactions with confidence by swiping, dipping, or tapping, which enables the quick pace that numerous smart cities would then aim to provide. The various challenges discussed in the previous section are about designing and implementing an extension mechanism to existing languages and/or systems so that infrastructures can be designed for smart cities with a high degree of interoperability, configurability, adaptability, and evolvability, with the following characteristics:
\begin{itemize}
\item Extension: a mechanism must be unified by generic integration that will enable different languages and system implementations to work together. 
\item Explicit models: clear definition of quality attributes has to be in place for interoperability, configurability, adaptability, and evolvability and the proposed extension mechanisms must be justified accordingly.
\item Declarative extension mechanisms over imperative ones.
\end{itemize}

Different areas of software engineering need to be explored and expanded in order to meet the new difficulties in order to handle some of these issues. This will frequently necessitate a total rethinking of the methodologies and the creation of new techniques, approaches, and instruments [57]. Achieving this requires the following elements: (1) further software engineering development to increase the potential for code reuse, reduce maintenance costs, improve usability, and integrate new functionality; New tools, methods, and metrics for evaluating the internal and external quality of software to drive development. Adapt to changing conditions. (2) new software architectures, architectural patterns, and methodologies to deal with the complexity of such systems; (3) brand-new tools, IDEs, and procedures along with instructions to encourage joint development of large-scale systems and applications emphasizing the importance of open source software communities.

\section{Conclusions}

The concept of the "Smart City" is rapidly becoming a reality, thanks to the integration of information and communication technologies, as well as cloud-based and mobile infrastructural facilities. Public organizations in numerous nations are utilizing these innovations to improve efficiency, sustainability, and the quality of life for their citizens. However, the heterogeneity of modern systems has made software development even more challenging, and it is now essential to recognize practices that are more appropriate given the technologies and background.

As the field of smart cities continues to develop, it faces fresh conceptual, technological, and academic obstacles. In this work, we discuss the top software engineering goals and challenges for smart city implementations. We found that suggested agile practices, such as continuous integration and daily meetings, were not being used to their full potential. Despite this category being instructed for the first time, we identified gaps that need to be addressed to determine which software engineering practices are more suitable for such a context.

Software engineering is essential to ensure that a connected digital platform drives the ambition for smart cities as it continues to take shape. To governmental bodies, small companies, and major corporations, software developers must produce comprehensive and coherent packages influenced by the most cutting-edge technological solutions. The concept of the "Smart City" is an exciting development that promises to revolutionize the way we live, work, and interact with our environment. However, it requires a concerted effort from all stakeholders to ensure that the software engineering practices employed are appropriate and effective. By doing so, we can create a sustainable and efficient future for our cities and citizens.

\bibliographystyle{unsrt}  
\bibliography{references}

\end{document}